# Scaling green hydrogen and CCUS via cement-methanol co-production in China

Yuezhang He[1,2,3], Hongxi Luo[2], Yuancheng Lin[1], Carl J. Talsma[4], Anna Li[3], Zhenqian Wang[1,5], Yujuan Fang[6], Pei Liu[1,*], Jesse D. Jenkins[2,3,*], Eric Larson[2,*], Zheng Li[1,6,*]

1. State Key Lab of Power Systems, Department of Energy and Power Engineering, Tsinghua University, Beijing, China
2. Andlinger Center for Energy and the Environment, Princeton University, Princeton, NJ, USA
3. Department of Mechanical and Aerospace Engineering, Princeton University, Princeton, NJ, USA
4. Carbon Solutions LLC, Bloomington, IN 47401
5. Research Institute of Petroleum Processing, Sinopec Corporation, Beijing, China
6. Laboratory of Low Carbon Energy, Tsinghua University, Beijing, China

## Abstract

High costs of green hydrogen and of carbon capture, utilization, and sequestration (CCUS) have hindered policy ambition and slowed real-world deployment, despite their importance for decarbonizing hard-to-abate sectors, including cement and methanol. Given the economic challenges of adopting CCUS in cement and green hydrogen in methanol production separately, we propose a renewable-powered co-production system that couples electrolytic hydrogen and CCUS through molecule exchange. We optimize system configurations using an hourly-resolved, process-based model incorporating operational flexibility, and explore integrated strategies for plant-level deployment and $CO_2$ source-sink matching across China. We find that co-production could reduce $CO_2$ abatement costs to $41–53 per tonne by 2035, significantly lower than approximately $75 for standalone cement CCUS and over $120 for standalone renewable-based methanol. Co-production is preferentially deployed at cement plants in renewable-rich regions, potentially reshaping national $CO_2$ infrastructure planning. This hydrogen–CCUS coupling paradigm could accelerate industrial decarbonization and scaling for other applications.

## 1. Main

Delivering on global climate targets requires rapid and far-reaching emissions reductions across all sectors, including heavy industries[1]. Green hydrogen and CCUS are widely regarded as indispensable for decarbonizing sectors such as steel, cement, and chemicals, where fossil feedstocks, high-temperature heat, and process-related carbon dioxide ($CO_2$) emissions limit the effectiveness of electrification and efficiency improvements[2,3]. In response, governments have announced increasingly ambitious policy targets, such as the EU's goal of producing 10 Mt of renewable hydrogen and building 50 Mt of $CO_2$ injection capacity annually by 2030[4,5]. However, real-world progress falls short of expectations and remains far from what net-zero pathways require due to high costs, infrastructure gaps, and uncertain market incentives[6,7].

Identifying and scaling early applications of electrolytic hydrogen and CCUS that are both economically viable and replicable can accelerate technological learning and enable broader adoption. One promising approach is to exploit industrial symbiosis, where distinct industrial processes are linked by exchange of mass and/or energy for mutual benefit[8]. Existing forms can improve resource efficiency, lower costs, and reduce emissions, as illustrated by cogeneration of heat and power[9], utilization of petroleum coke for aluminum anodes[10] and substitution of cement clinker with steel slag[11]. However, many of these have limited potential because they serve niche markets or operate only in highly specialized contexts. Symbiotic coupling of electrolytic hydrogen and CCUS through molecular exchange and process integration, as proposed here, could unlock novel and scalable applications of these decarbonization-enabling technologies. The oxygen ($O_2$) co-produced from water electrolysis can supply oxy-fuel combustion CCUS systems (e.g. high-temperature cement kilns)[12], thereby increasing flue-gas $CO_2$ concentration and reducing or even eliminating the need for air separation units (ASU). Meanwhile, the captured $CO_2$ can serve as a carbon feedstock, which together with electrolytic hydrogen, can be used to synthesize methanol and other hydrocarbon-based chemicals, reducing or eliminating $CO_2$ transport and sequestration costs. In contrast to the predominantly one-way material or energy flows in most industrial symbiosis, coupling green hydrogen and CCUS could create bidirectional molecular linkages. Symbiotic coupling of cement and methanol production integrated with green hydrogen and CCUS is proposed here as a strategy for decarbonizing those industries while also supporting commercial development of green hydrogen and CCUS technologies.

In 2022, cement and methanol production globally emitted approximately 2.4 Gt $CO_2$[13] and 0.26 Gt $CO_2$[14], respectively, together accounting for around 7% of energy related emissions globally. China accounted for 57% of cement and 37% of the methanol production (2020)[15,16], with a large share of carbon-intensive coal-based routes. For cement, efficiency improvements, clinker substitution, and fuel switching can reduce fossil energy use, but roughly two thirds of emissions are from carbonate calcination and essentially unavoidable without capture, which makes CCUS indispensable[17]. For methanol, dependence on fossil derived hydrogen and carbon feedstocks means that deep reductions require cleaner inputs, such as renewable hydrogen and carbon neutral $CO_2$ from biogenic sources or direct air capture[18,19]. China is already piloting cement CCUS using oxy-fuel combustion[20] and renewable hydrogen-based methanol projects that source $CO_2$ from biomass or industrial flue gas[21,22]. However, standalone deployment of CCUS in cement and renewable hydrogen-based methanol remain expensive. Equipment level improvements in cement capture systems and plant level source–sink optimization can reduce capture, transport and storage costs[23,24]. Renewable methanol pathways that pair electrolytic hydrogen with direct air capture and operational flexibility can lower levelized costs[25,26]. Yet, the green premium is estimated at over 40% for cement and more than 100% for methanol compared to conventional pathways[26,27].

Linking cement and methanol, two historically unconnected sectors, through green hydrogen and CCUS could reduce green premiums for both. Previous studies on methanol production explored clinker kiln off-gases as a carbon source, including using captured $CO_2$ for catalytic hydrogenation with renewable hydrogen[19], employing electrochemical $CO_2$ reduction[28], and reforming flue gases with natural gas[29]. For cement, prior work on cement oxy-fuel combustion evaluated the benefits of utilizing byproduct $O_2$ from water electrolysis to replace $O_2$ from ASU[30]. Our previous work proposed a co-located configuration that enables bidirectional exchange of $O_2$ and $CO_2$ between cement kilns and methanol synthesis units[31]. We found synergistic economic and environmental benefits for both sectors and estimated that such co-production routes could account for 24% of

China's clinker output and 55% of its methanol production by 2060. Despite these promising findings, two critical aspects of the co-production system warrant deeper investigation.

One is flexibility of operations. Traditionally, industrial facilities have been designed for continuous operation, relying on stable fossil energy supplies to maximize utilization and economic performance. Integrating variable renewable energy introduces variability and intermittency that require storage and flexible operation across units and time scales[32,33]. In the co-production system, multiple process streams and interlinked energy and material flows create a complex, tightly coupled multi-vector topology, calling for a unified optimization modeling framework that captures cross commodity flexibility and interdependent storage dynamics.

The second is facility-level deployment and system-level planning. The economic viability of co-production hinges on renewable quality and on proximity to $CO_2$ sequestration sites. Spatial heterogeneity in wind, solar and sequestration availability thus mandates a two-tiered approach: plant-specific techno-economic analysis using high-resolution renewable profiles, coupled with system-level geospatial $CO_2$ pipeline and storage network optimization to identify least-cost clusters for co-production roll-out.

In the present work, we develop a detailed process-based and hourly-resolved model that incorporates unit level flexibility under solar and wind supply and explicitly manages flows of $H_2$, $O_2$ and $CO_2$, as shown in Fig. 1. We represent the entire supply and utilization chain of commodities within the co-production system of cement fueled by biomass and methanol synthesized from green $H_2$ and captured $CO_2$, covering production, storage (in different states), consumption, and sequestration (if applicable), as shown in Extended Data Fig. 1. Then, we optimize capacity sizing, flexible operation, and storage across different cement-to-methanol stoichiometries (see Table 1). We conduct plant-level analysis of existing cement plants to evaluate the cost-effectiveness and spatial heterogeneity of technology deployment. We integrate the process optimization model with a SimCCS-China model[34] to optimize $CO_2$ source-sink matching considering costs of pipeline construction across China. This enables generating comprehensive deployment strategies of co-production technologies under different $CO_2$ sequestration scenarios. We find that this co-production system could achieve a much lower overall abatement cost than decarbonizing these two sectors separately, primarily benefiting from more economical sources of $O_2$ and $CO_2$. Unlocking the flexibility of methanol synthesis and configuring molecular storage buffers offer additional but different by plant economic benefits compared to steady and continuous production. Additionally, introducing the co-production technology could have a significant impact on $CO_2$ transport network planning in China's cement industry, by prioritizing CCS implementation at cement facilities with abundant proximate renewable resources.

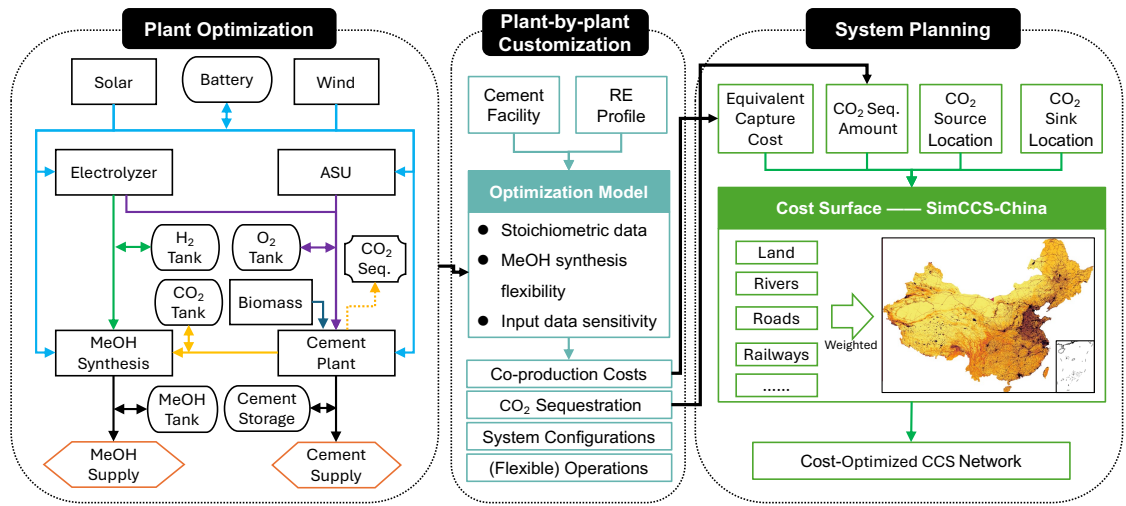

**Fig. 1: Schematic diagram of the methodological framework.**

| Scenario | $CO_2$ Handling Strategy | Cement-to-Methanol Stoichiometry | $CO_2$ emission reduction (%) | Methanol Synthesis Flexibility | Notes |
|---|---|---|---|---|---|
| Net-Zero Cement & Methanol | Partial $CO_2$ sequestration allowed | 9.76 | 100% | Flexible | Aligned with certification schemes (e.g., ISCC Sustainable Marine Fuels) that require methanol to use biogenic or atmospheric $CO_2$ for long-term shipping fuel eligibility. |
| | | | | Inflexible | |
| No-$CO_2$ Sequestration | All captured $CO_2$ used for methanol synthesis | 2.77 | 84% | Flexible | Represents a closed-loop carbon pathway; no geological sequestration is allowed. |
| | | | | Inflexible | |
| Separate Decarbonization | Cement and methanol decarbonized independently | 9.76 | 77% | Flexible | Cement with standalone CCUS; methanol produced via standalone green $H_2$ and $CO_2$ sourced from DAC. |
| | | 2.77 | 89% | | |

**Table 1: Scenario settings for cement–methanol system decarbonization.** $CO_2$ emission reduction is defined as the proportion of $CO_2$ emissions avoided relative to incumbent coal-based technologies, specifically for the case where methanol is used as a shipping fuel. The accounting scope includes direct $CO_2$ emissions from cement and methanol production, indirect $CO_2$ emissions from grid electricity, and direct $CO_2$ emissions from methanol combustion, assuming complete combustion (see details in Section 3 in SI).

**Co-production reduces $CO_2$ abatement costs by lowering supply cost of $O_2$ and $CO_2$**

We find that the co-production system can achieve a 38%-65% reduction in $CO_2$ abatement costs compared to decarbonizing the two plants separately for a selected case-study facility in Inner Mongolia (**Fig. 2**a). Despite slightly higher H2 costs (**Fig. 2**b), the cost savings primarily stem from lower-cost $O_2$ (**Fig. 2**c) and $CO_2$ (**Fig. 2**d). The increase in H$_2$ costs arises because $O_2$ consumption and $CO_2$ generation in cement kiln introduce inflexible elements to the system, limiting the ability of $H_2$ production to follow renewable availability as flexibly as in standalone $H_2$-based methanol. By diminishing the ASU, the system lowers both capital investment and energy consumption, resulting in a 71%-95% decrease in $O_2$ costs (**Fig. 2**c). However, due to the temporal mismatch between variable $O_2$ production and fixed $O_2$ consumption, $O_2$ storage and its associated processes constitute the largest share of the $O_2$ cost structure. In the no-$CO_2$ sequestration scenario, the amount of byproduct $O_2$ is 2.7 times the $O_2$ demand, eliminating the necessity for $O_2$ storage tanks. Capturing, processing and storing $CO_2$ from cement production is 80%-91% cheaper than costs of $CO_2$ from DAC, as shown in **Fig. 2**d. We use a relatively optimistic DAC cost at $203.6/t $CO_2$[26], within the typical $200–600/t $CO_2$ range (see Section 3 in SI for the literature review on DAC). Applying partial $CO_2$ sequestration could reduce $CO_2$ abatement costs by further leveraging the economically sourced $O_2$.

Also, unlocking the flexible operation of methanol synthesis could reduce overall production costs by 8%-12% primarily due to lower $H_2$ costs, despite slight increases in $O_2$ and $CO_2$ costs. Harnessing demand-side flexibility in $H_2$ consumption can reduce the levelized cost of electricity (LCOE) from solar and wind sources and the required $H_2$ storage tank size, ultimately lowering the levelized cost of hydrogen (LCOH) by 22%-24% (**Fig. 2**b). Inflexible methanol synthesis requires a stable, hour-by-hour supply of $CO_2$ and $H_2$. The fixed $CO_2$ extraction from clinker kilns aligns perfectly with the stable $CO_2$ demand, eliminating the need for $CO_2$ storage tanks. However, with flexible methanol synthesis scenarios, the flexible $CO_2$ demand and stable $CO_2$ supply create operational asynchrony and necessitate $CO_2$ storage tanks as a buffer. Similarly, $O_2$ supply is more compatible with a fixed demand, which reduces the cost of $O_2$ through downsized storage tanks and a smaller complementary ASU. Thus, maximizing the flexibility options could reduce overall system costs, but it requires more coordinated and complex system design and operations.

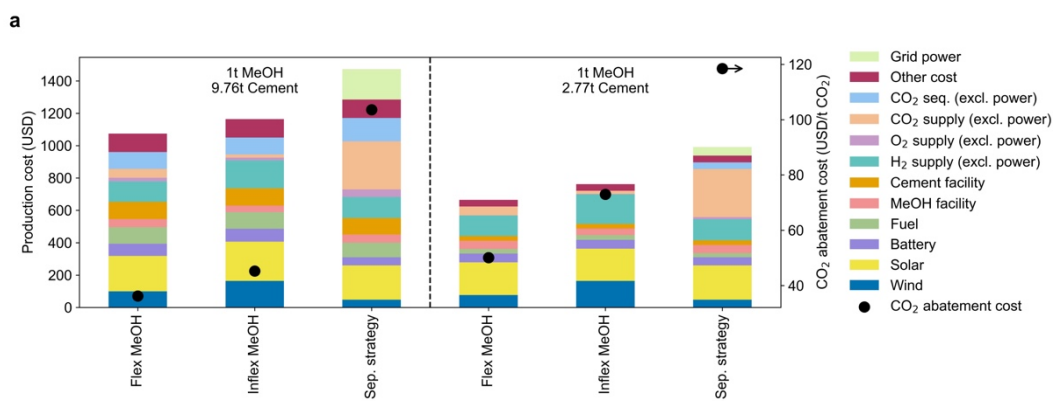

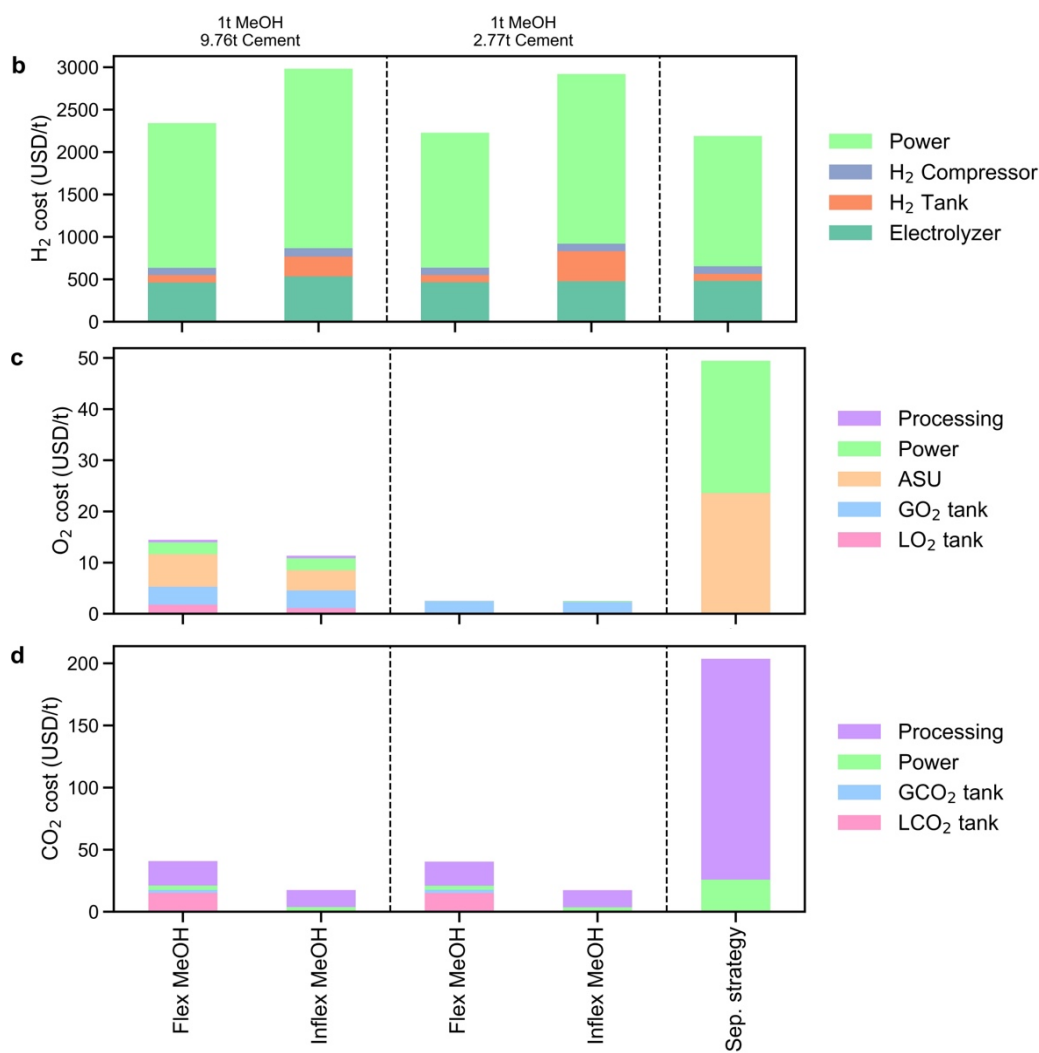

**Fig. 2: Cost structure of the co-production system and associated molecules for the selected plant in Inner Mongolia. a**, Cost breakdown of the co-production technology. $CO_2$ abatement cost is defined as the cost difference between the decarbonized technology and the incumbent dominant route without emissions abatement (i.e., coal-based production and current grid electricity), divided by the corresponding emission reduction. In this case study, the $CO_2$ abatement cost for separate decarbonization is estimated at $79 per tonne of $CO_2$ for cement and $131 per tonne of $CO_2$ for methanol (see Section 3 in SI for detailed calculations). The $CO_2$ abatement cost presented in this figure reflects the combined cost for cement and methanol under a given stoichiometric ratio. "Other costs" include labor and miscellaneous cement raw materials. "$CO_2$ seq." represents the costs associated with $CO_2$ compression, transport and sequestration, assuming the transport cost is $8.7/ t $CO_2$ and storage cost is $5.8/ t $CO_2$[35]. **b**, Cost breakdown of $H_2$ supply. **c**, Cost breakdown of $O_2$ supply. **d**, Cost breakdown of $CO_2$ supply. "Processing" in **c** and **d** refers to capital and operational costs associated with compressors (for gaseous storage) and liquefaction units (for liquid storage).

## Optimized system configurations and flexible operations

By coupling flexibilities across processes, the variability of renewable generation cascades through all production and storage stages, with each component contributing to accommodating renewable fluctuations. We observe distinct patterns of optimal flexible operations across components within the co-production system under the Net-Zero Cement & Methanol scenario (**Fig. 3**). In the case-study plant in Inner Mongolia, electrolyzer loads closely follow the combined generation of solar and wind, operating primarily during the daytime while shutting down or reducing output at night. Methanol synthesis exhibits seasonal variations, operating at full load during most of the spring. Batteries, exhibiting an intra-day cycle, predominantly charge during the daytime and discharge at night. $H_2$ and gaseous $O_2$ storage tanks possess similar annual charge-discharge cycles, typically ranging from hours to days. Gaseous $CO_2$ tanks cycle 21 times per year, but with uneven temporal distribution across the year. In contrast, liquid $O_2$, liquid $CO_2$, and methanol storage function as long-duration storage, with the number of annual round-trip cycles ranging from 2.1 to 3.9. While liquid $O_2$ and methanol are charged in spring and summer when renewable generation is abundant and discharged in other periods, liquid $CO_2$ follows the opposite pattern.

The discrepancy in storage behavior among different commodities can be explained by unit capital cost considerations. Storage technologies with low capital costs per unit of stored volume (e.g., liquid storage) tend to exhibit long-duration storage, despite the high capital and energy penalties associated with charging and discharging, aligning with findings in a previous study[36]. Conversely, storage with higher volume capacity costs (e.g., gaseous storage) is inclined to increase cycling frequency, diluting capital costs across higher turnover rates to improve cost efficiency per unit of stored commodity or energy. Overall, these operational patterns reflect the underlying variability of renewable resources, highlighting that the value of flexibility lies in enabling reliable integration of intermittent solar and wind.

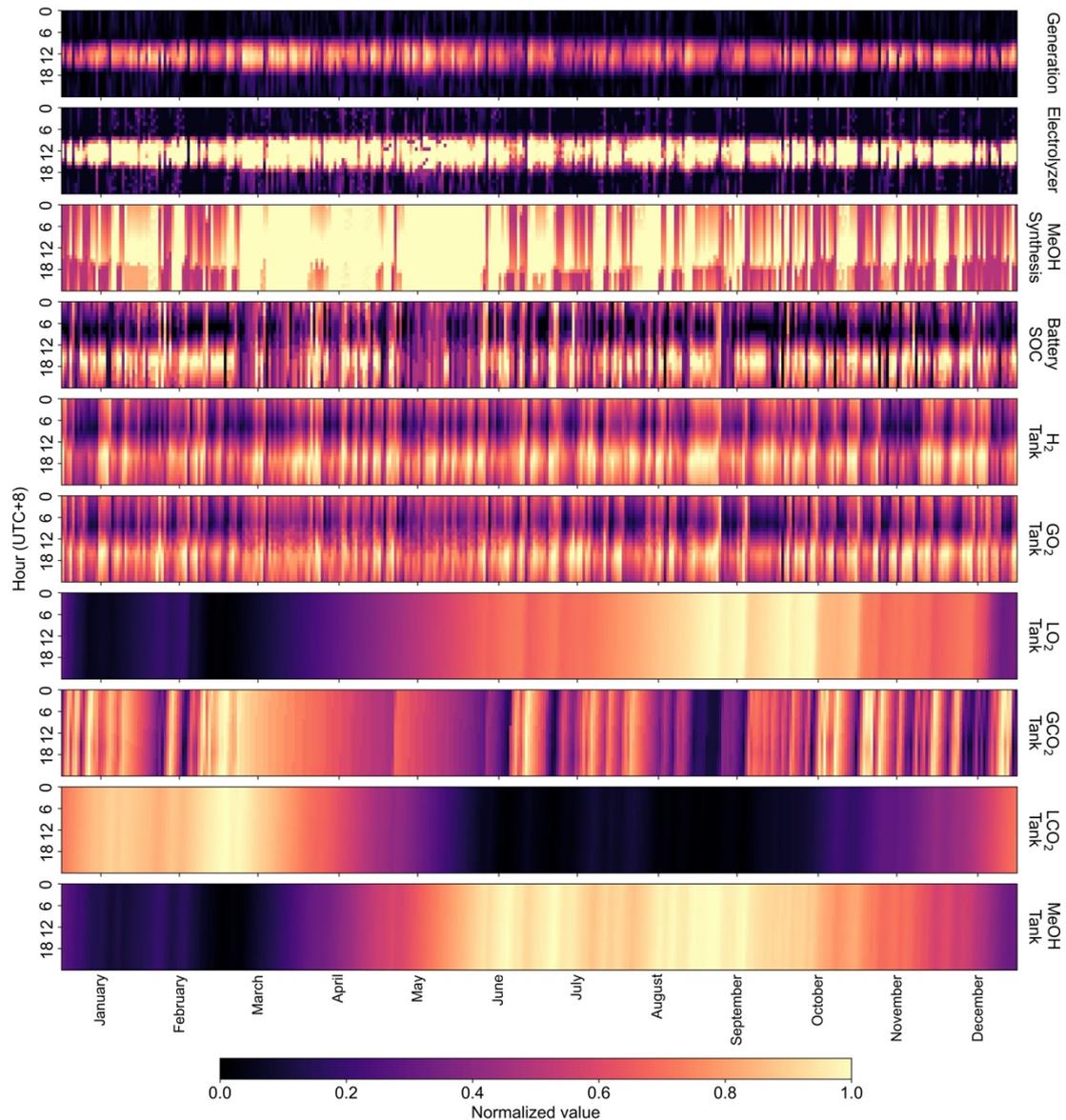

**Fig. 3: Annual normalized operational profiles of the selected plant in Inner Mongolia under the Net-Zero Cement & Methanol scenario with flexible methanol synthesis.** The x-axis represents 365 days over one year, and the heat maps illustrate hourly variations, with the y-axis indicating 24 hours per day. Values are normalized by dividing hourly absolute values by their annual maximum (linear normalization). "Generation" denotes the actual hourly power generation from solar and wind, normalized by the weather-dependent maximum available hourly generation. "Electrolyzer" and "MeOH synthesis" indicate hourly loads normalized by their respective maximum loads. "State of charge (SOC)" for Battery, $H_2$ tank, gaseous $O_2$ ($GO_2$) tank, liquid $O_2$ ($LO_2$) tank, gaseous $CO_2$ ($GCO_2$) tank, liquid $CO_2$ ($LCO_2$) tank, and MeOH tank refer to the hourly SOC normalized by each component's maximum SOC over the entire year. The operations under inflexible methanol synthesis are shown in Extended Data Fig. 2.

In contrast to RE-rich weeks characterized by nearly full-load methanol synthesis, RE-scarce weeks feature frequent ramping, which drives distinct patterns in $H_2$, $O_2$, and $CO_2$ production, storage, and consumption. We analyze representative RE-scarce and RE-abundant weeks for the plant in Inner Mongolia, focusing on the balance of electricity, $H_2$, $O_2$, and $CO_2$, as shown in **Fig.**

4. Electrolyzers account for 85% of power consumption annually, with batteries discharging primarily at night to maintain the minimum electrolyzer load (5%). During RE-rich weeks, electrolyzers could operate overnight, powered by strong wind generation. $H_2$ storage is discharged at night to supplement the minimum $H_2$ supply required for methanol synthesis. Beyond real-time consumption, most excess $O_2$ is stored in gaseous form, supplementing demand at night. During RE-scarce weeks, liquid $O_2$ storage experiences lower charging and higher discharge, while ASU operates at a higher load compared to RE-rich weeks. Meanwhile, some $O_2$ may be vented in RE-rich weeks, indicating a trade-off between maximizing byproduct $O_2$ utilization and minimizing $O_2$ storage system costs. For $CO_2$ storage, gaseous $CO_2$ remains largely idle, while liquid $CO_2$ undergoes net discharge during RE-abundant weeks. Conversely, in RE-scarce weeks, gaseous $CO_2$ is charged at night and utilized on days with relatively higher renewable output, whereas liquid $CO_2$ is charged continuously without discharge.

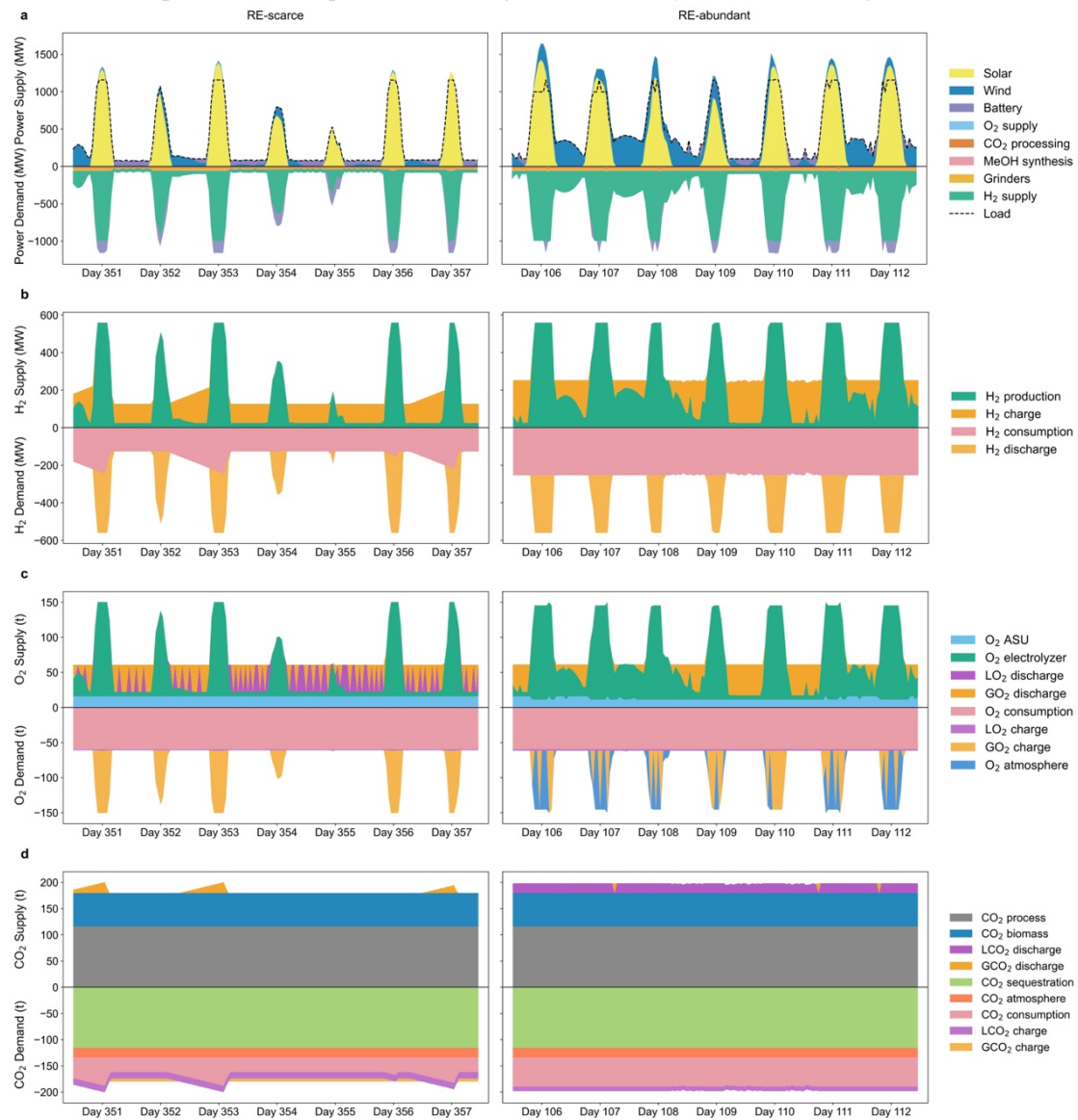

**Fig. 4: Operational profiles under renewable energy (RE)-scarce (left panel) and RE-abundant (right panel) weeks for the selected plant in Inner Mongolia under the Net-Zero Cement & Methanol scenario with flexible methanol synthesis.** Positive values indicate the supply (including discharge) of each commodity, while negative values indicate demand

(including charge). **a**, Electricity balance. The dashed line indicates the aggregated electricity load; power generation exceeding the load is curtailed. "$O_2$ supply" includes electricity consumed by the ASU, compressors, and liquefaction units. "$CO_2$ processing" includes electricity consumed by compressors and liquefaction units. "$H_2$ supply" represents electricity consumption for water electrolysis and $H_2$ compression. **b**, $H_2$ balance. **c**, $O_2$ balance; "$O_2$ atmosphere" indicates vented $O_2$. **d**, $CO_2$ balance; "$CO_2$ process" indicates $CO_2$ from carbonate decomposition, "$CO_2$ biogenic" indicates $CO_2$ from biomass combustion, and "$CO_2$ atmosphere" refers to $CO_2$ emitted to the atmosphere due to less than 100% capture rate. The operations under inflexible methanol synthesis are shown in Extended Data Fig. 3.

**Geographic differences in plant economic performance due to RE spatial heterogeneity**

Given the estimated shrinking cement demand[37,38] and the expanding methanol market[39] in China, we consider a scenario where new methanol facilities are co-located with existing cement plants (823 plants in total). We focus on greenfield methanol facilities because the current methanol fleet is predominantly coal-based and incompatible with low-carbon integration, whereas cement could involve retrofit. In the Net-Zero Cement & Methanol scenario (9.76 tonnes cement per tonne methanol), the spatial heterogeneity of renewable resources leads to variations in $CO_2$ abatement costs (see **Fig. 5**a, excluding $CO_2$ transport and sequestration costs) and results in diverse optimized system configurations and operational strategies (see Section 1 in SI). Cement plants located in Northeast China and North-Central China, where renewable resources are abundant, achieve lower $CO_2$ abatement costs (excluding $CO_2$ transport and storage), approximately $15–35 per metric ton. For example, the case plant in Inner Mongolia falls within this range, with an abatement cost of $31 per metric ton. In contrast, plants located in central China exhibit higher abatement costs, often exceeding $40 per metric ton of $CO_2$.

Unlocking the flexibility of methanol synthesis yields plant-specific cost reductions, with cement plants in China achieving co-production cost savings ranging from 3% to 17% (**Fig. 5**b). Generally, plants with lower baseline co-production costs under inflexible methanol synthesis exhibit smaller marginal benefits from adding flexibility. Nevertheless, enabling flexible methanol synthesis can still achieve over $5 per metric ton in $CO_2$ abatement cost savings for each unit of added capacity following the ascending order of abatement cost (**Fig. 5**c).

Although $CO_2$ abatement costs for most plants are more sensitive to solar costs, the lowest-cost plants rely more heavily on high-quality wind resources and are therefore more sensitive to wind costs. For most plants, solar power accounts for a larger share of total system costs, averaging 11% compared with 5% for wind and 4% for electrolyzers. However, for plants with relatively low $CO_2$ abatement costs, wind cost parameters can emerge as the dominant sensitivity factor. For example, among the 50 lowest-cost plants, wind averages 9% of total costs compared with 6% for solar. These results suggest that while large-scale cost reductions depend on declining solar costs, achieving the lowest abatement costs might require prioritizing deployment in regions with high-quality wind resources.

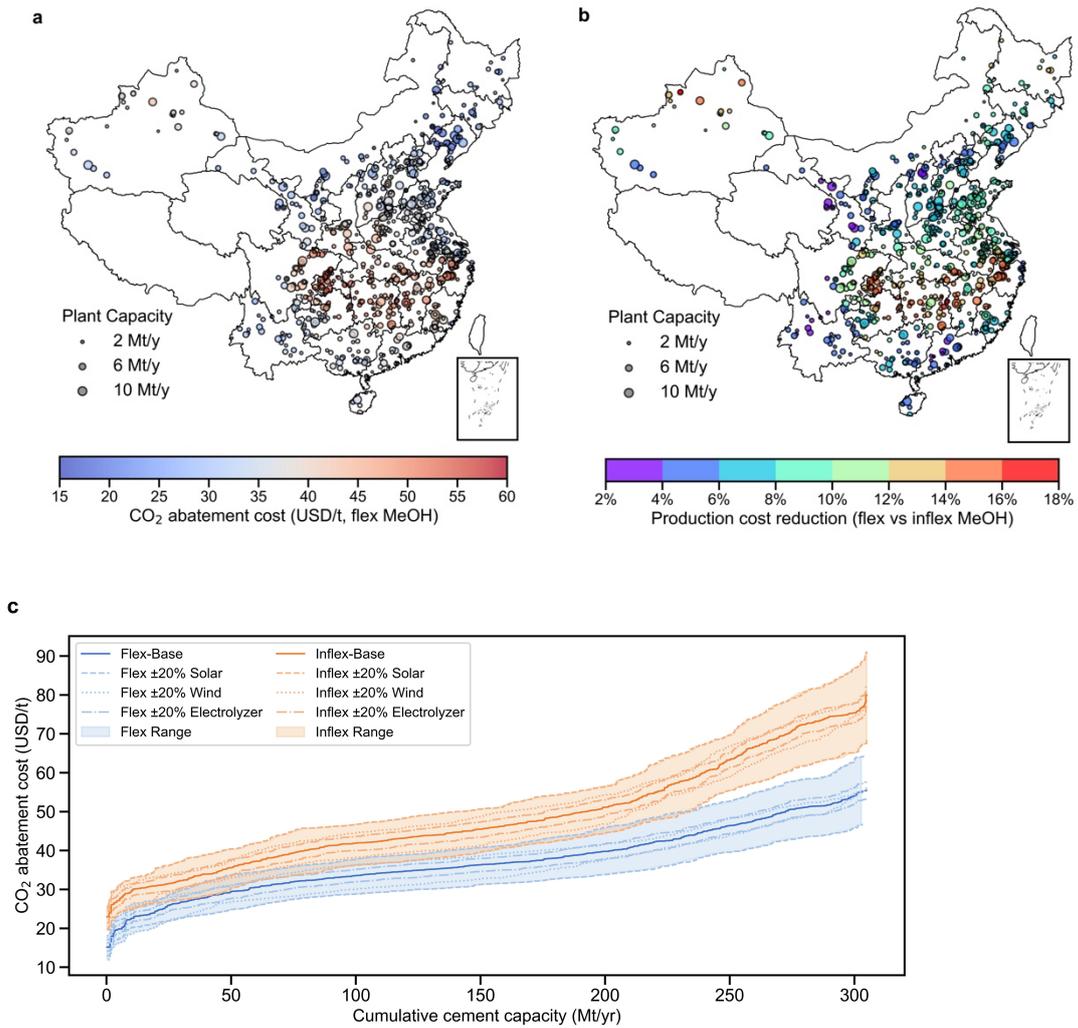

**Fig. 5: Plant-by-plant economic performance and sensitivity analysis in China under Net-Zero Cement & Methanol scenarios. a**, $CO_2$ abatement cost by plant, defined as $(cost_{decarbonized} - cost_{incumbent})/\Delta CO_2$ per 9.76 t cement + 1 t MeOH. Incumbent routes use coal-based production and current grid electricity; $CO_2$ transport and sequestration costs are excluded. Each dot marks an existing cement plant with different capacity (size) and abatement cost (color). **b**, the ratio of cost difference between flexible (Flex) and inflexible (Inflex) methanol synthesis, on the same 9.76 t cement + 1 t MeOH basis. **c**, cost–capacity curves sorted in ascending order of abatement cost versus cumulative cement capacity. The solid blue line is the Flex baseline; lighter blue lines show sensitivity to renewable-related cost parameters under Flex scenarios; the light blue shaded band is the Flex envelope (min–max). The sensitivity analysis is conducted by changing the cost parameters of solar, wind and electrolyzer by 20% one at a time. The solid orange line is the Inflex baseline; lighter orange lines show sensitivity of Inflex; the light orange shaded band is the Inflex envelope. It is assumed that local biomass resources are sufficient to meet the demand under the Net-Zero Cement & Methanol scenario, with supporting estimates and an availability analysis provided in Section 4 in SI.

**Effect of co-production on planning of $CO_2$ source-sink matching**

Although the abovementioned optimization model evaluates the economic performance based on renewable resource availability, incorporating post-capture (i.e., transportation and storage) costs may alter the overall production cost distribution and influence the optimal plant selection strategy. Consequently, a co-optimization approach that integrates production and post-capture costs provides insights into the most cost-effective deployment of co-production facilities. To address this, we employ an extended version of SimCCS-China to co-optimize onsite production costs, $CO_2$ pipeline network design, and $CO_2$ storage costs under different $CO_2$ sequestration amount scenarios (see Methods and Section 6 in SI for details). As shown in **Fig. 6**a and **Fig. 6**c, the selected cement plants largely align with plants that have lower production costs shown in **Fig. 5**a. The "$CO_2$ sequestration 26.0 Mt/yr" scenario corresponds to 7.06 Mt/yr of methanol output under co-production, while the "$CO_2$ sequestration 311.5 Mt/yr" scenario corresponds to 84.7 Mt/yr. At the initial-deployment stage, the low-hanging fruit for adopting co-production technologies is found in Northeast China, where the additional costs from $CO_2$ transport and sequestration are approximately $15 per tonne. As $CO_2$ sequestration increases to large-scale deployment, $CO_2$ captured from some cement plants in North China, Northeast China, Southwest China, and East China is transported to nearby sequestration sites through a $CO_2$ pipeline network. However, compared to the plant distribution shown in **Fig. 5**a, some plants with low production costs are not connected to the network due to their long distances to suitable storage sites.

Meanwhile, the introduction of the co-production system significantly alters the $CO_2$ source-sink matching layout compared to decarbonizing China's cement sector alone (**Fig. 6**b and **Fig. 6**d). Under a cement-only decarbonization strategy, $CO_2$ pipelines connect fewer plants located in renewable-rich regions such as North China and Northeast China, but with potential expansions into central China. $CO_2$ transport costs are higher when using co-production technologies, driven by lower capturable $CO_2$ volumes per plant and more pipeline installations required (**Fig. 6**e). This indicates that producing net-zero methanol through co-production systems involves a cost trade-off, where increased $CO_2$ transport requirements are offset by lower production-side abatement costs.

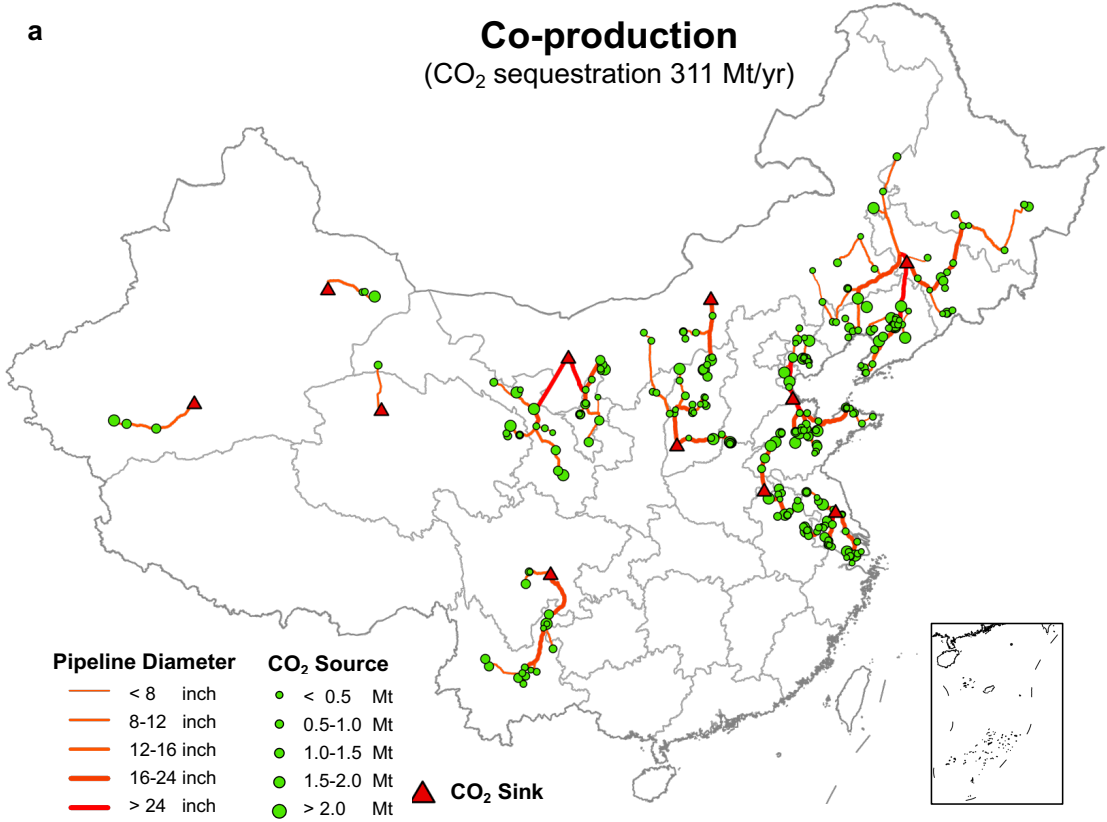

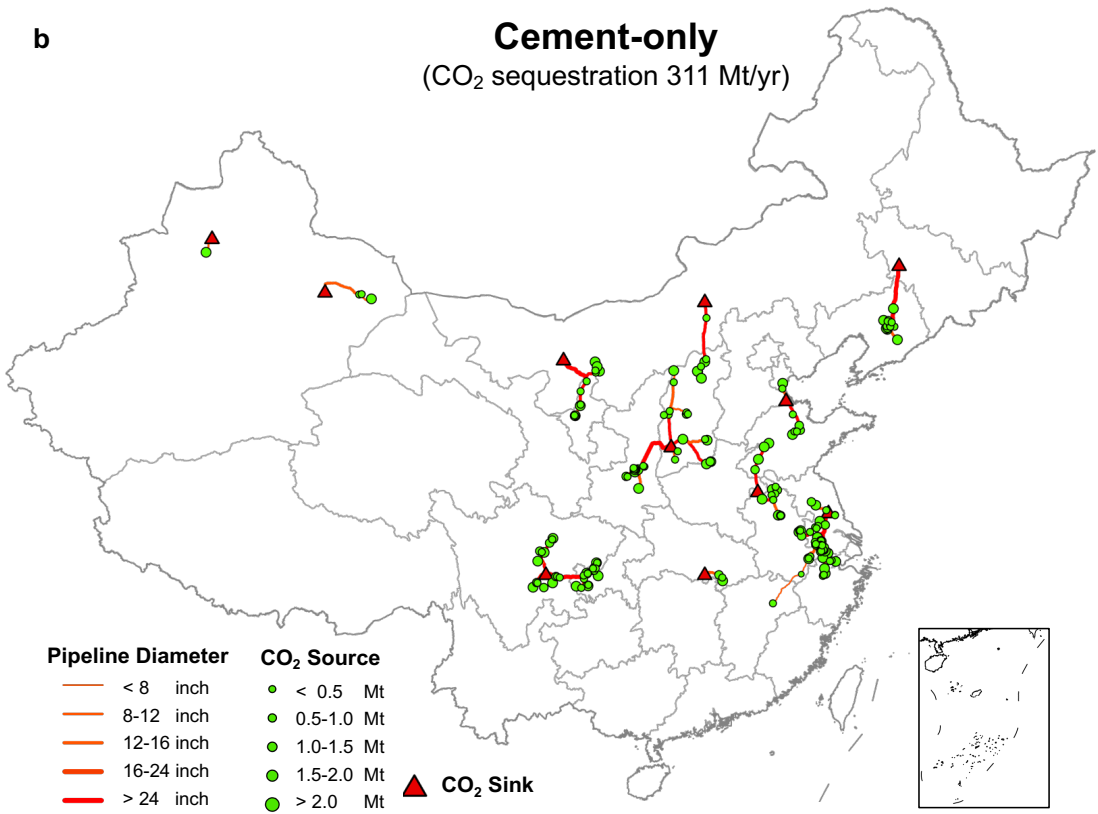

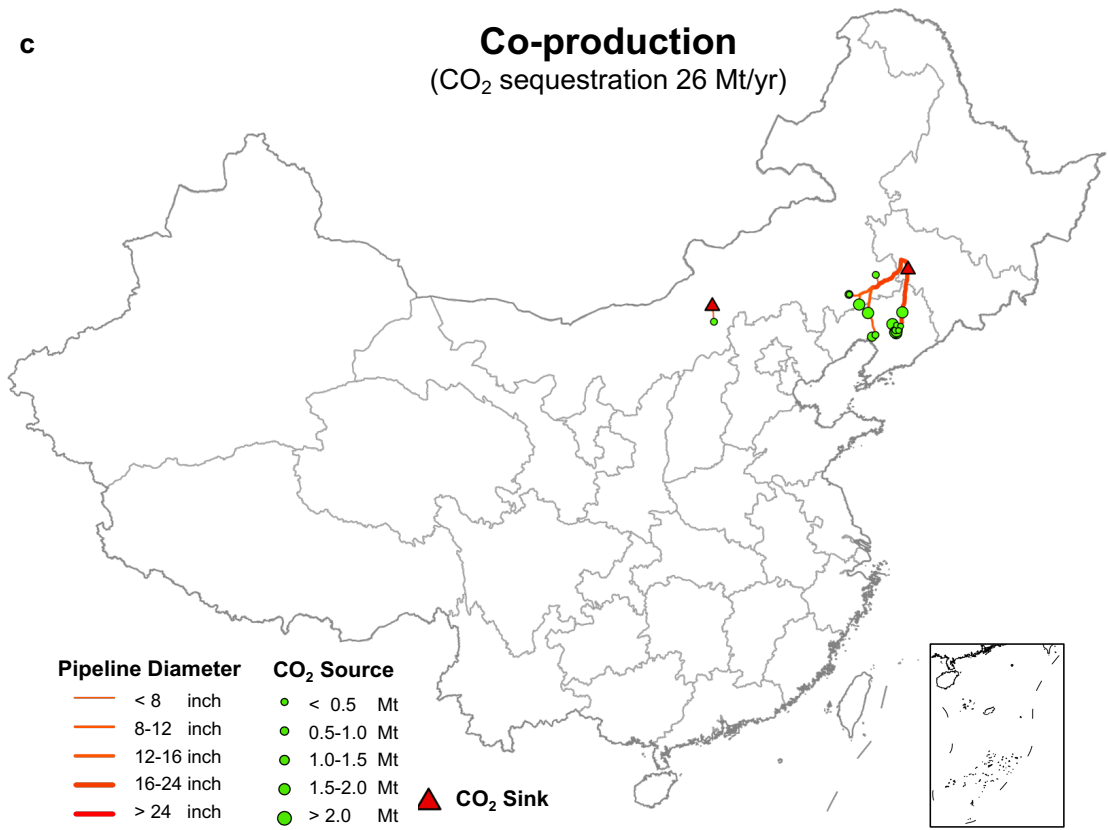

**c** **Co-production**
($CO_2$ sequestration 26 Mt/yr)

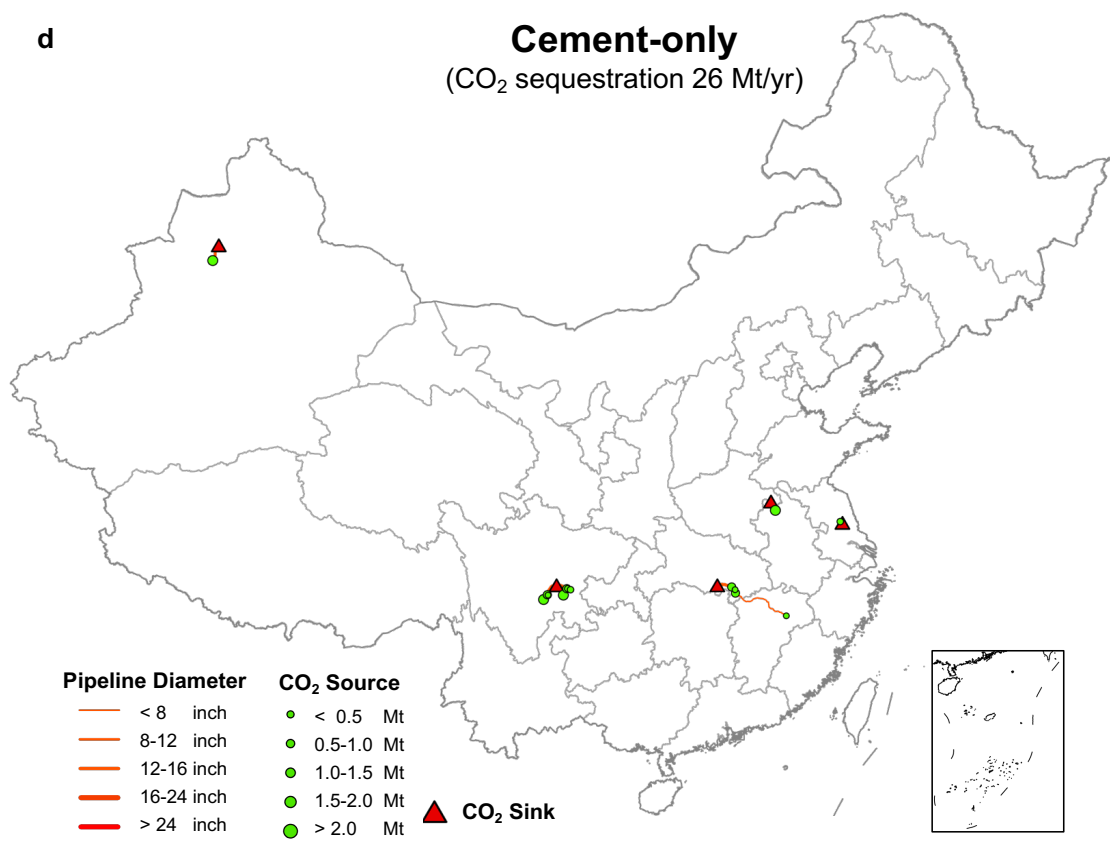

**d** **Cement-only**
($CO_2$ sequestration 26 Mt/yr)

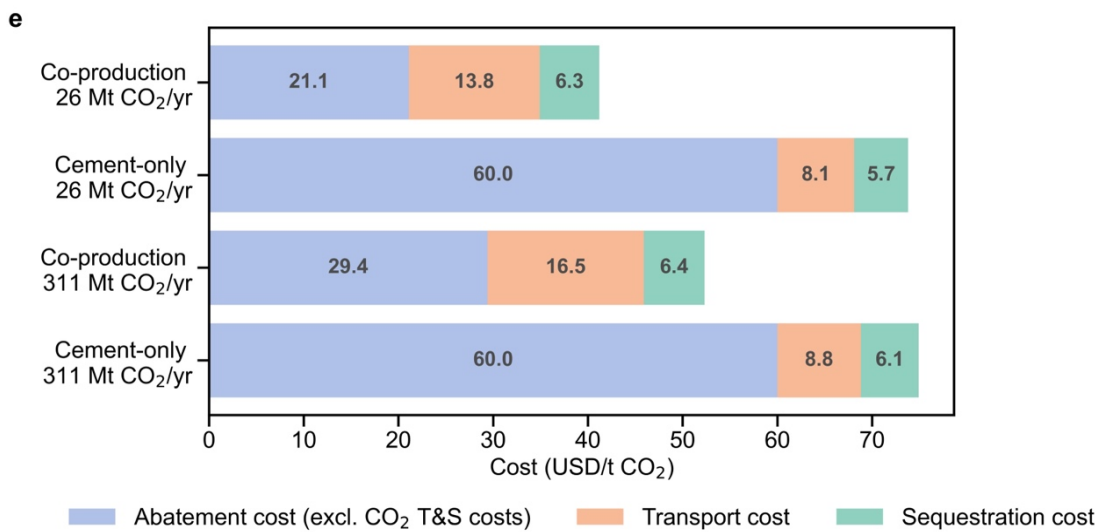

**Fig. 6: Comparison of $CO_2$ source–sink matching under co-production versus cement-only decarbonization scenarios.** Compared to the case-study plant in Inner Mongolia, here we change the $CO_2$ transport and sequestration costs from fixed value per tonne of $CO_2$ into variable numbers (see Methods). "$CO_2$ sequestration 311 Mt/yr" refers to the scenario where all incumbent methanol production (70.6 Mt) and an estimated share of marine fuel demand (14.1 Mt) are substituted with green methanol under co-production. "$CO_2$ sequestration 26 Mt/yr" corresponds to a scenario with a 10% substitution of incumbent methanol production (7.06 Mt) by co-production. Subfigures **a**–**d** show four combinations of $CO_2$ sequestration amount and technology adoption. Green circles mark $CO_2$ point sources at cement plants, with size representing capturable emission volume; red triangles mark $CO_2$ storage sinks. Orange lines denote the proposed $CO_2$ pipeline network with different diameters reflected by line width. Subfigure **e** shows the $CO_2$ abatement cost structure (see Method for calculations). Additional $CO_2$ sequestration scenarios are presented in Section 7 in SI.

## 2. Discussion

Co-production, as proposed in this study provides substantial economic advantages over separate decarbonization of the cement and methanol sectors. By 2035, co-producing net-zero cement and methanol in China could lower $CO_2$ abatement costs to $41–53 per tonne, markedly below approximately $75 for independent CCS in cement and over $120 for independent renewable-based methanol. With China's carbon price projected to reach $30–70/t$CO_2$ by 2035[40], the co-production approach holds promise to achieve cost parity with conventional technologies under plausible policy scenarios while achieving co-decarbonization. Spanning from detailed process modeling to system-level planning, this study offers critical insights into cost-effective and feasible decarbonization pathway for these two hard-to-abate sectors through optimized co-production systems. However, successful deployment will require strong financial incentives, targeted policy support for demonstration projects, and coordinated corporate strategies bridging these traditionally unrelated industries. The limitations of this study are listed in section 8 in SI.

Leveraging industrial symbiosis requires reevaluating the role and carbon footprint of industrial feedstock molecules when integrating nascent decarbonization technologies. $H_2$ has attracted extensive attention for its potential as industrial feedstock[41], spurring technoeconomic analyses and lifecycle emission assessments[42,43]. Conversely, the application potential of renewable-sourced $O_2$ and $CO_2$ as industrial feedstocks remains underexplored. Currently, industrial $O_2$ is primarily sourced through energy-intensive air separation for combustion, chemical synthesis, and metallurgy[44]. In a future industrial system deploying large-scale electrolytic hydrogen, high-purity $O_2$ generated from water electrolysis could be directed for oxy-fuel combustion, improving carbon capture efficiency and enhancing system-wide decarbonization economics by reducing costs associated with air separation[45]. Meanwhile, captured $CO_2$ can serve as a raw material for chemical synthesis where a carbon source is needed[46]. Using captured $CO_2$ is significantly more economical than direct air capture and less emission-intensive than conventional coal-based routes.

Such sector-coupled industrial symbiosis, enabled by the exchange of byproduct $O_2$ and captured $CO_2$, can be extended to other emerging industrial applications. Industries that rely on hydrogen and $CO_2$ as feedstocks, such as synthetic fuels for aviation and shipping, can benefit from upstream sources of high-purity $CO_2$. Meanwhile, sectors such as thermal power, glass, and lime production involve high-temperature combustion processes. Applying oxy-fuel combustion in these industries can generate high-concentration $CO_2$ emissions, while the associated $O_2$ demand can be partially met by the $O_2$ co-produced from electrolytic hydrogen systems. Thus, the generalized co-production paradigm represents an innovative form of industrial symbiosis that integrates oxy-fuel combustion–CCUS with electrolytic hydrogen, opening new frontiers industrial decarbonization and delivering mutual benefits for traditionally unconnected sectors.

Exploring the flexibility of industrial processes may further lower the costs of low-carbon technologies. A unique feature of the renewable-based co-production system is its integration of operational flexibility across different components and storage methods across multiple molecules. Under the flexible methanol synthesis scenario, the temporal mismatch of supply and demand is characterized by: (1) $H_2$: flexible production and flexible consumption; (2) $O_2$: flexible production and fixed consumption; (3) $CO_2$: fixed production and flexible consumption. Different storage methods (e.g., liquid vs. gaseous storage) operate at varying storage durations, shaped by their economic and technical parameters. The temporal asynchrony of molecular flows arising from renewable energy variability underscores the necessity and complexity of strategically coordinating operational modes and deploying molecular buffering strategies to optimize system-wide adaptability. Therefore, designing such a system, which deviates from stable, continuously operated industrial facilities, requires new modeling tools. In practice, despite considerable theoretical cost savings, implementing flexible operations poses challenges related to operational feasibility and increased O&M costs, requiring further industrial demonstration to validate their real-world applicability.

Implementing co-production technologies further demands customized deployment strategies to address the spatial heterogeneity of renewable resources, $CO_2$ pipeline costs, and sequestration site distributions. Our results indicate that optimal system configurations, economic performance, and potential savings from flexibility vary plant by plant, leading to plant-specific optimal operations across different components. Additionally, achieving net-zero emissions requires a co-optimization of renewable-based production alongside $CO_2$ transport and sequestration planning. Compared to cement-only decarbonization, adopting co-production technologies results in different $CO_2$ pipeline layouts, both for initial-deployment and large-scale $CO_2$ sequestration scenarios. Given the substantial investments needed for $CO_2$ pipeline infrastructure, the deployment of co-production technologies must carefully assess the economic viability and technical feasibility of $CO_2$ network planning, and be coordinated with potential pipelines designed for the cement sector alone.

Introducing co-production technologies could also influence the decommissioning strategy for China's cement production capacity. In 2020, China produced 2395 Mt cement[47]; however, demand has begun declining, driven by a weakened real estate and infrastructure market, and is projected to decrease by 42-79% by 2060[37,38]. Due to overcapacity and future demand decline, the criteria guiding decisions on cement plant retention are likely to prioritize factors such as plant efficiency rankings, availability of surrounding decarbonization resources (e.g., $CO_2$ sequestration sites, alternative fuels), and regional cement demand[24,31,48]. Co-production technologies might confer strategic advantages to cement plants located in regions with abundant renewable resources or even stimulate the construction of new capacity in these regions through capacity replacement policies. In contrast, China's methanol market remains robust, supported by both conventional chemical demand and emerging applications such as low-carbon fuels[49]. Our analysis suggests that under the scenario of substituting all conventional methanol and methanol shipping fuel forecast in 2050, co-production technologies would necessitate utilizing approximately 41.3% of China's current cement capacity (see Section 7 in SI). Therefore, dynamic planning of cement plant decommissioning and the incremental co-production deployment should be comprehensively evaluated for optimizing China's long-term decarbonization pathway.

## 3. Methods

**Optimization model**

The process-detailed linear programming optimization model is designed to optimize capacity sizing, flexible operations, and storage strategies for each system component at an hourly resolution under constraints (see Section 2 in SI). Electricity is sourced from a combination of dedicated local solar and wind sources. Locally-sources biomass is used for clinker calcination, ensuring a carbon-neutral energy supply during the cement production stages. Alkaline electrolyzers generate $H_2$ and $O_2$ simultaneously with a flexible load. $H_2$ must be compressed to 150 bar to meet the pressure requirements for chemical production before being used in methanol synthesis or stored in gaseous $H_2$ tanks. $O_2$ can be utilized immediately, stored in gaseous form, or liquefied for long-duration storage. Since the operating pressure of clinker kilns (~1 bar) is lower than the electrolyzer outlet pressure, $O_2$ can be used without compression. To manage temporal mismatches, $O_2$ can either undergo compression and be stored in gaseous tanks (lower energy penalty but higher capital cost per mass stored), or be liquefied and stored in liquid tanks (higher energy penalty but lower capital cost per mass stored). While byproduct $O_2$ utilization for oxy-fuel combustion offers economic benefits, an ASU is also included to supplement $O_2$ during temporal shortages or when elevating the stoichiometric balance between cement and methanol reduces byproduct $O_2$ availability. Cement production is divided into three stages: raw meal preparation, clinker production, and cement production. The output commodity of each stage can be easily stored since the solid outputs have simple storage requirements. Raw meal preparation and cement production involve grinding processes powered by renewable electricity and can operate flexibly. However, the clinker kiln must operate steadily due to the stability requirements of chemical reactions at more than 1500°C[50]. $CO_2$ emissions from clinker kilns originate from biomass combustion (~1/3) and carbonate decomposition (~2/3). Captured $CO_2$ can be stored in liquid or gaseous form for future use, or compressed and transported for underground sequestration. For immediate consumption or gaseous storage, $CO_2$ is compressed to 50 bar, while for sequestration, it is compressed to 150 bar. Similar to $O_2$ storage, $CO_2$ storage options include gaseous and liquid storage. Methanol is synthesized from green $H_2$ and captured $CO_2$ and stored in liquid form. Methanol synthesis possesses the potential for flexible operation within a certain load range and ramping limits. We assume a fixed hourly supply for both cement and methanol. The resulting optimization problem involves hundreds of thousands of variables and constraints.

**Input data**

Technical and economic parameters were derived from authors' process simulations and data from the literature. We use ASPEN Plus to simulate the detailed production processes for methanol synthesis, and $O_2$/$CO_2$ compression and liquefaction systems (Section 5 in SI). The energy consumption for these processes is extracted from the simulations. Stoichiometric data for other processes are obtained from the literature and summarized in our previous work (see Table S2 in SI). Flexibility-related parameters, including ramp rates and adjustable load ranges across all components, are also sourced from the literature (see Table S4 in SI). The costs of renewable generation technologies, batteries, and electrolyzers are customized for China using 2035 projections. Economic data for other facilities are obtained from the literature (see Table S3 in SI). Renewable energy profiles are sourced from Renewable.ninja, using 2019 weather data. The biomass cost is set at $80 per tonne (dry matter). Costs are reported in 2020 USD.

**Stoichiometric balance**

The stoichiometric balance between cement and methanol is defined as the hourly supply mass ratio of these two products. The minimum stoichiometric value occurs when all $CO_2$ from clinker production is directed to methanol synthesis without $CO_2$ sequestration. As the stoichiometric number increases, higher cement production corresponds to greater $O_2$ generation and $CO_2$ utilization. Excess $CO_2$ is sequestered, and additional $O_2$ is from otherwise-vented electrolytic $O_2$ and/or increased $O_2$ from air separation. A greater proportion of sequestered $CO_2$ also improves the overall $CO_2$ removal rate. In the net-zero scenario, the amount of biogenic $CO_2$ that is sequestered equals the uncaptured $CO_2$ emissions plus the $CO_2$ utilized in methanol synthesis, which is ultimately released into the atmosphere. We evaluate various stoichiometric configurations and present the corresponding economic and environmental performance in Extended Data Fig. 4.

**Plant-level customization**

We obtain plant-level cement facility data[52] and retrieve hourly renewable energy profiles for each plant location. The national average cement-to-clinker ratio is used to estimate the clinker kiln size at each facility. Given that the typical production line capacity ranges from 2,500 to 10,000 metric tons of clinker per day, we exclude outliers because erroneously large values exceed the capacity of any existing cement plant in China, while very small plants are unlikely to be prioritized for CCUS deployment. The total selected cement plant capacity is 2,976 Mt per year, corresponding to an average capacity utilization rate of 80% in 2020. We assume this capacity utilization rate remains constant in this study to convert capacity size into cement production. Emission factors and energy efficiency per unit output are kept consistent across all facilities. For each plant, we customize facility size and location-specific renewable profiles. These parameters are then used to run the process-detailed optimization model to assess economic performance, commodity output, and $CO_2$ emissions. At this stage, each plant is optimized independently without considering interactions with other facilities.

**$CO_2$ source-sink matching**

Achieving net-zero cement and methanol production requires sequestering a portion of the captured $CO_2$. This necessitates building $CO_2$ transport and storage infrastructure, and calls for an optimization-based approach to match $CO_2$ sources with nearby sequestration sinks to minimize system costs. Cement plants are widely distributed because the low-value-added cement products are highly sensitive to transportation costs. Suitable $CO_2$ storage sites, primarily saline aquifers, are concentrated in specific regions. This results in varying $CO_2$ transport distances for different plants. $CO_2$ can be transported via trucks, ships, or pipelines. This study adopts pipeline transportation due to its advantages in enabling large-scale, long-distance $CO_2$ transport for onshore applications and relatively low operational emissions.

**Input data preparation**: The construction cost of $CO_2$ pipelines varies across regions due to geographic, topographic, and demographic factors. Our previous study on $CO_2$ source-sink matching in China's steel sector developed a geographic information system (GIS)-based cost surface with a 240×240 m resolution for pipeline construction[34]. In this study, we update the geographical coordinates of $CO_2$ sinks from prior research by incorporating new basins (see Section 6 in SI). For $CO_2$ sources, we primarily evaluate the net-zero scenario by adopting the co-production strategy and compare it with the cement decarbonization-only scenario. The cement decarbonization-only scenario involves implementing CCS technology at selected cement plants without additional decarbonization measures (e.g., fuel switching, efficiency improvements), while achieving the same level of $CO_2$ sequestration as the net-zero scenario. Under the net-zero scenario, we derive plant-level $CO_2$ emissions and co-production costs (excluding $CO_2$ transport and storage) from the process-detailed optimization model. In the cement decarbonization-only scenario, capturable $CO_2$ emissions at the plant level are assumed to be proportional to plant capacity, with cement production costs per tonne remaining constant (see Section 7 in SI).

SimCCS is a decision-support software designed for the optimization and integration of CCUS technologies. One version of SimCCS generates the optimal pipeline layout under specified $CO_2$ sequestration targets. The input data includes geographical coordinates of $CO_2$ sinks and sources, the capturable $CO_2$ amount and capture costs by source, the sequestrable $CO_2$ amount and sequestration costs by sink, and the pipeline construction cost surface. The objective function minimizes the whole process cost, including capture, transport infrastructure, and sequestration costs, as defined in Eq. (1), where $i$ denotes the $CO_2$ source adopting CCS, and $j$ represents the $CO_2$ sink utilized.

In this study, we extend the SimCCS-China framework to optimize the overall $CO_2$ processing costs (equivalent to minimizing the $CO_2$ abatement cost) under predefined targets. The objective function in Eq. (2) replaces the capture cost in Eq. (1) with the equivalent capture cost, which is derived by converting the co-production cost into an equivalent cost per unit of $CO_2$ sequestered (Eq. (3)). At the optimum, the co-production cost, minus the cost of the coal-based benchmark technology and divided by the amount of $CO_2$ avoided, yields the system-level $CO_2$ abatement cost. We set different scales of green methanol output and convert these targets into $CO_2$ sequestration requirements based on the stoichiometric relationship between utilized and sequestered $CO_2$. Without considering $CO_2$ transport and sequestration costs, the co-production system is naturally favored in regions with abundant renewable energy. We incorporate post-capture (i.e., transportation and storage) costs into the optimization to evaluate their impact on cost distribution and plant siting, and perform a geospatial assessment of $CO_2$ sequestration resources to enable source–sink matching for net-zero methanol and cement production.

$$Min \sum_{i,j} (Capture\_cost_i + Pipeline\_cost_{i,j} + Sequestration\_cost_j) \quad (1)$$

$$Min \sum_{i,j} (Eq.capture\_cost_i + Pipeline\_cost_{i,j} + Sequestration\_cost_j) \quad (2)$$

$$\begin{aligned} Eq.capture\_cost(\$/tCO_2) &= co\\&-production\ cost(\$/(t\ MeOH + xt\ cement))/stoichiometry(x)\\&/CO_2\ sequestration\ factor\ (tCO_2/t\ cement) \end{aligned} \quad (3)$$

## Acknowledgments

The authors gratefully acknowledge the support provided by Saudi Aramco-Tsinghua University cooperative project, Princeton University's Carbon Mitigation Initiative and Andlinger Center for Energy and the Environment.

## Author contributions

Y.H., H.L., P.L. and J.D.J. conceptualized the study and methodology. Y.H. H.L. and C.T. implemented the software, validated the model and analysed the data, with contributions from Y.L. and Z.W. on data curation and visualization. Y.H. and Y.L. wrote the original draft, with major revisions from E.L., A.L., P.L., J.D.J., H.L., Y.F. and Z.L.. E.L. and Z.L. acquired the funding. Z.L., P.L., E.L. and J.D.J. supervised the study.

## Declaration of interests

J.D.J. serves on the advisory boards of Eavor Technologies Inc., a closed-loop geothermal technology company, Rondo Energy, a provider of high-temperature thermal energy storage and industrial decarbonization solutions, Dig Energy, a developer of low-cost drilling solutions for ground-source heat pumps, and Karman Industries, a developer of advanced heat pumps for industrial applications and has an equity interest in each company. He also serves as a technical advisor to MUUS Climate Partners and Energy Impact Partners, both investors in early-stage climate technology companies.

# Extended data

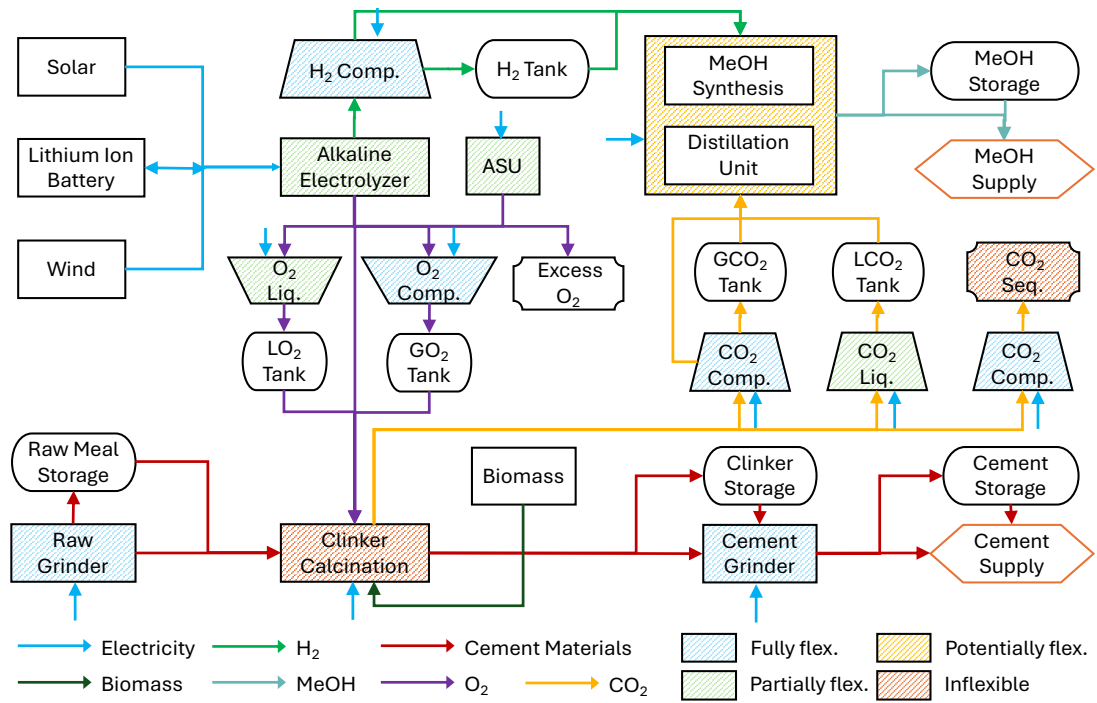

**Extended Data Fig. 1 Framework of the co-production system.** The system is powered by solar, wind, and biomass, with all components optimized for capacity sizing and hourly operations. Storage systems for electricity, $H_2$, $O_2$, $CO_2$, methanol, and cement-related materials are modeled to support asynchronous operations and balance the temporal mismatch of supply and demand. Four levels of operational flexibility are defined: fully flexible (rapid response across full load with current technologies), partially flexible (response within a limited load range or subject to ramping constraints), potentially flexible (limited flexibility under current technologies but with future potential), and inflexible (constrained by inherent technical limitations). Cement and methanol supply are exogenously fixed on an hourly basis.

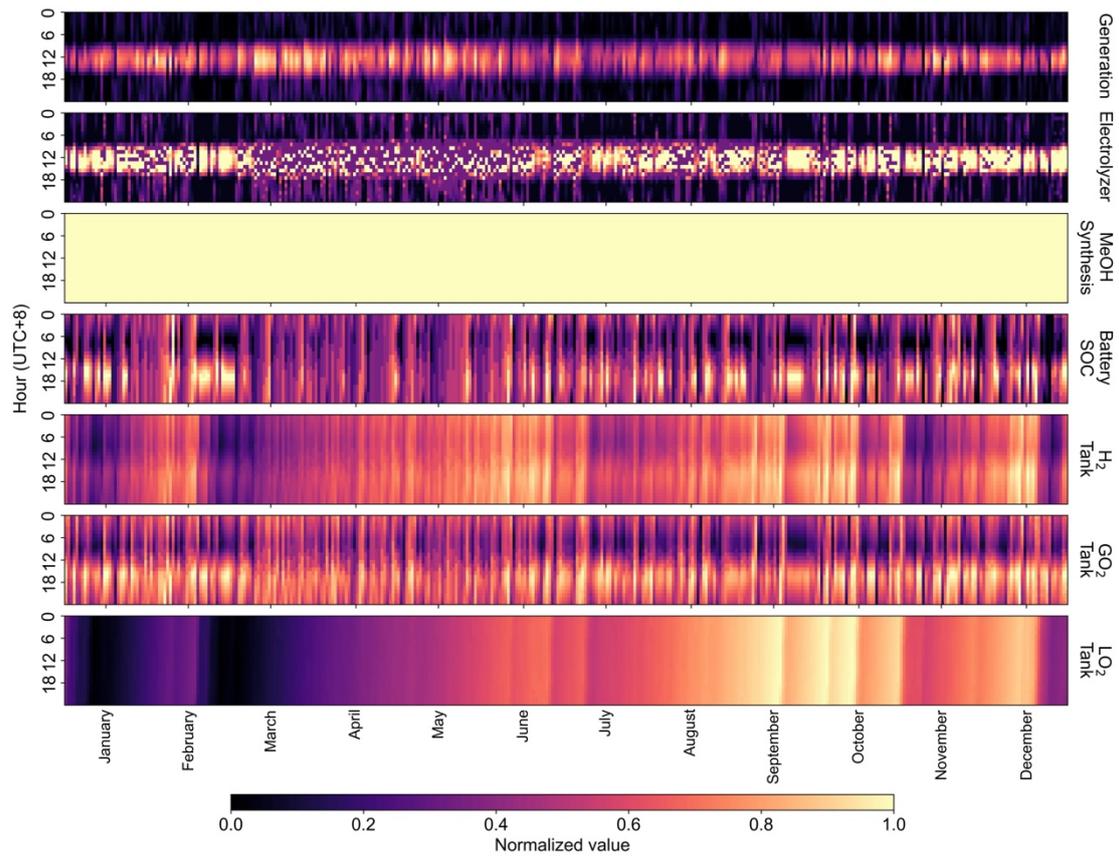

**Extended Data Fig. 2 Annual normalized operational profiles of the selected plant in Inner Mongolia under the Net-Zero Cement & Methanol and inflexible methanol synthesis scenario.** The x-axis represents 365 days over one year, and the heat maps illustrate hourly variations, with the y-axis indicating 24 hours per day. Values are normalized by dividing hourly absolute values by their annual maximum (linear normalization). "Generation" denotes the actual hourly power generation from solar and wind, normalized by the weather-dependent maximum available hourly generation. "Electrolyzer" and "MeOH synthesis" indicate hourly loads normalized by their respective maximum loads. "State of charge (SOC)" for Battery, $H_2$ tank, gaseous $O_2$ ($GO_2$) tank, and liquid $O_2$ ($LO_2$) tank refers to the hourly SOC normalized by each component's maximum SOC over the entire year.

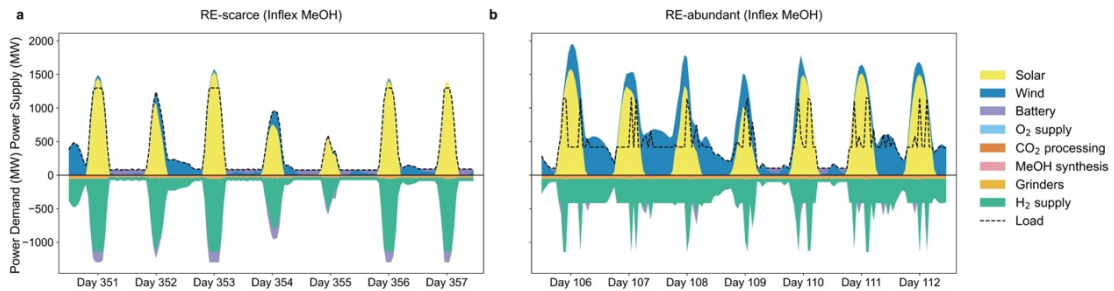

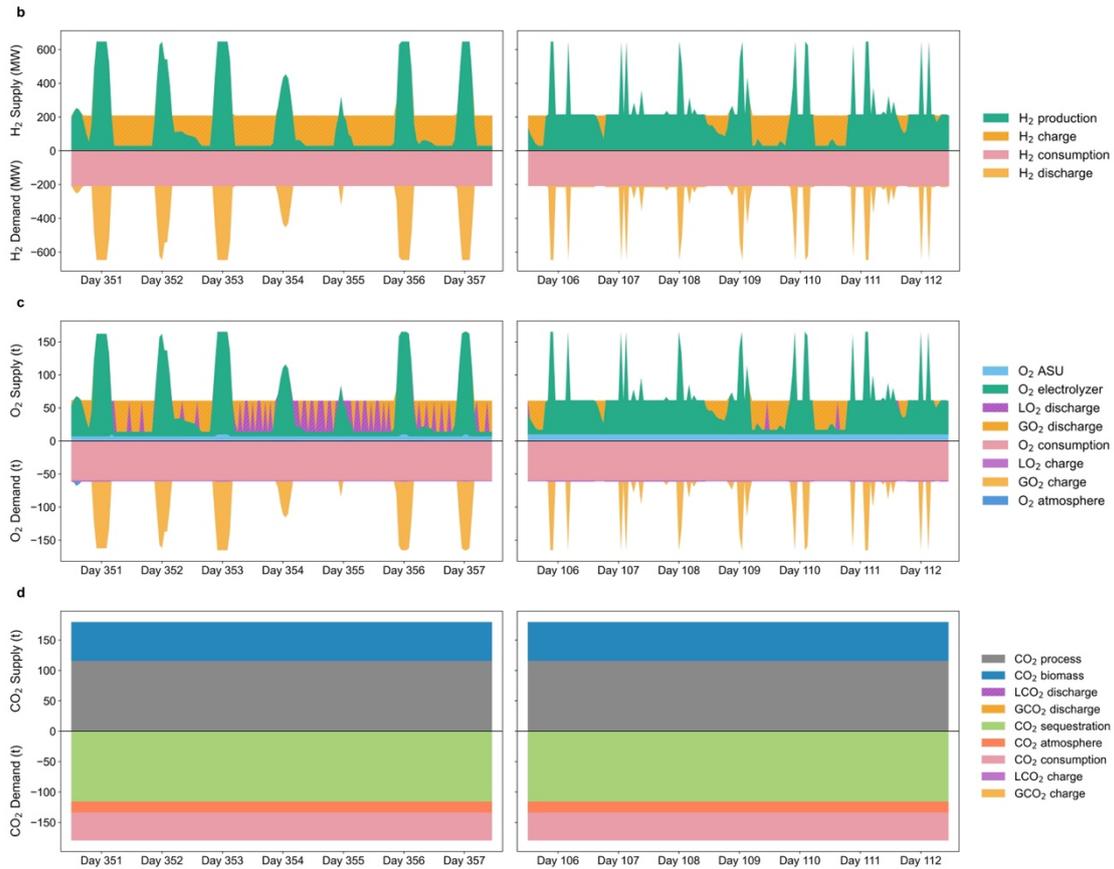

**Extended Data Fig. 3 Operational profiles under renewable energy (RE)-scarce and RE-abundant weeks for the selected plant in Inner Mongolia under the Net-Zero Cement & Methanol and inflexible methanol synthesis scenario.** The left panels illustrate operations during a representative RE-scarce week, while the right panels show a representative RE-abundant week. In each subfigure, positive values (above the x-axis) indicate the supply (including discharge) of each commodity, while negative values (below the x-axis) indicate demand (including charge). **a**, Electricity balance. The dashed line indicates the aggregated electricity load; power generation exceeding the load is curtailed. "O₂ supply" includes electricity consumed by the ASU, compressors, and liquefaction units. "CO₂ processing" includes electricity consumed by compressors and liquefaction units. "H₂ supply" represents electricity consumption for water electrolysis and H₂ compression. **b**, H₂ balance. **c**, O₂ balance; "O₂ atmosphere" indicates discarded O₂. **d**, CO₂ balance; "CO₂ process" indicates CO₂ from carbonate decomposition, "CO₂ biogenic" indicates CO₂ from biomass combustion, and "CO₂ atmosphere" refers to CO₂ emitted to the atmosphere due to less than 100% capture rate.

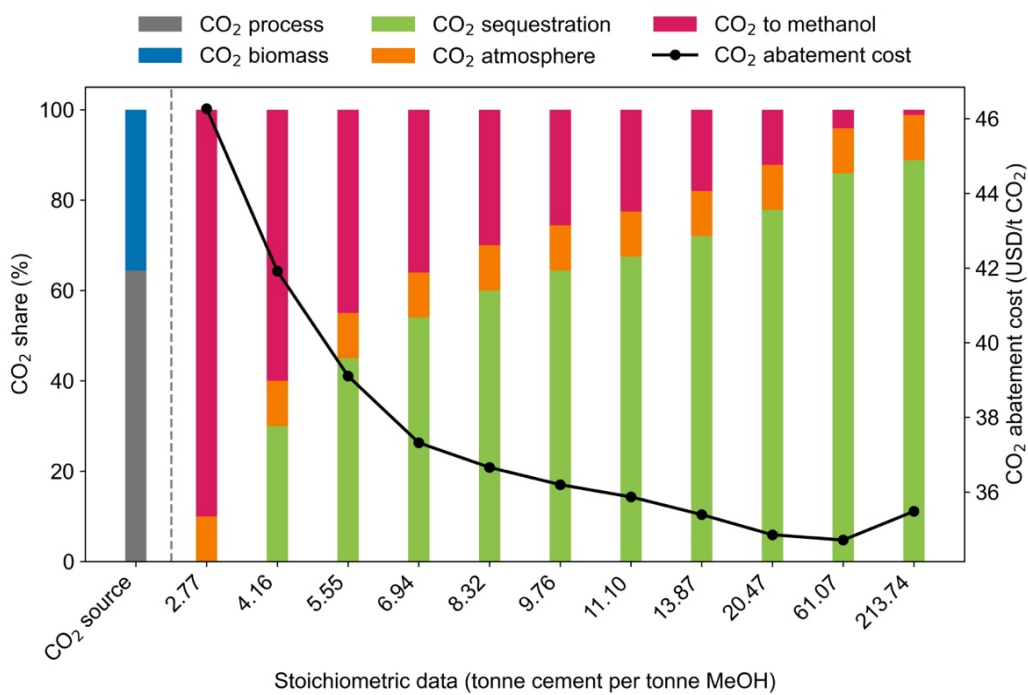

**Extended Data Fig. 4 CO₂ abatement cost and CO₂ share under different cement-to-methanol stoichiometric data.** The leftmost bar shows the composition of the CO₂ source—exclusively from cement facilities—which remains the same across all scenarios. The subsequent bars indicate the share of CO₂ delivered to each final destination for the stoichiometric datasets on the x-axis. The line with markers represents the CO₂ abatement cost, calculated assuming CO₂ transport and storage costs of $8.7 and $5.8 per tonne of CO₂, respectively. As the ratio increases, CO₂ sequestration rises while CO₂ used for methanol synthesis correspondingly falls. This shift drives the CO₂ abatement cost down from $46 /tCO₂ in the no-sequestration scenario to $36 /tCO₂ under the net-zero methanol scenario (stoichiometric ratio = 9.76). Although abatement costs are minimized at stoichiometric ratios between 20.5 and 213.7, the incremental reduction relative to the net-zero methanol scenario (ratio = 9.76) is limited.